\documentclass[aps,PRResearch,twocolumn,superscriptaddress,floatfix,amsmath,amssymb,showpacs,preprintnumbers]{revtex4-2}
\usepackage{amsmath,amsfonts,amssymb,amsthm,graphics,graphicx,epsfig,bbm}
\usepackage[colorlinks=true,citecolor=blue,linkcolor=blue,urlcolor=blue]{hyperref}
\usepackage[usenames]{color}
\usepackage[utf8]{inputenc}
\usepackage{graphicx}
\usepackage{subfigure}
\usepackage{amsmath}
\usepackage{epsfig}
\usepackage{dcolumn}
\usepackage{bm}
\usepackage{color}
\usepackage{times}
\usepackage{epstopdf}
\usepackage{amssymb}
\usepackage{lipsum}
\usepackage{amstext}
\usepackage{latexsym}
\usepackage{hyperref}
\usepackage{amsfonts}
\usepackage{psfrag}
\usepackage{dsfont}
\usepackage{braket}
\usepackage{comment}
\UseRawInputEncoding

\usepackage[nameinlink]{cleveref}
\crefname{equation}{Eq.}{}
\Crefname{equation}{Eqs.}{}
\crefname{figure}{Figs.}{Figs.}
\Crefname{figure}{Fig.}{Fig.}

\crefname{table}{Table}{Table}
\crefname{Appendix}{Appendix}{Appendices}

\begin{document}

\title{Active Carpets in floating viscous films}


\date{\today}

\author{Felipe A. Barros}
\email{fbarros2017@udec.cl}
\affiliation{Departamento de F\'isica, Facultad de Ciencias, Universidad de Concepci\'on, Concepci\'on, Chile}

\author{Hugo N. Ulloa}
\email{ulloa@sas.upenn.edu}
\affiliation{Department of Earth and Environmental Science, University of Pennsylvania, Philadelphia, USA}

\author{Gabriel Aguayo}
\email{g.aguayo07@gmail.com}
\affiliation{Departamento de F\'isica, Facultad de Ciencias F\'isicas y Matem\'aticas, Universidad de Chile, Santiago Chile}

\author{Arnold J. T. M. Mathijssen}
\email{amaths@sas.upenn.edu}
\affiliation{Department of Physics and Astronomy, University of Pennsylvania, Philadelphia, USA}

\author{Francisca Guzm\'an-Lastra}
\email{fguzman@uchile.cl}
\affiliation{Departamento de F\'isica, Facultad de Ciencias, Universidad de Chile, Santiago ,Chile}

\begin{abstract}
Earth's aquatic environments are inherently stratified layered systems where interfaces between layers serve as ecological niches for microbial swimmers, forming colonies known as \textit{Active Carpets}. Previous theoretical studies have explored the hydrodynamic fluctuations exerted by \textit{Active Carpets} in semi-infinite fluid media, demonstrating their capability to enhance thermal diffusion and mass transport in aquatic systems. Yet, little is understood about the fluid dynamics and impact of \textit{Active Carpets} residing in confined layered environments, like slicks floating on water bodies. In this study, we report novel solutions for the hydrodynamic fluctuations induced by \textit{Active Carpets} geometrically confined between a free surface and a fluid-fluid interface characterized by a jump in fluid viscosity. Combining theory and numerical experiments, we investigate the topology of the biogenic hydrodynamic fluctuations in a confined, thin fluid environment. We reveal that within this thin layer, \textit{Active Carpets} gives shape to three characteristic regions: Region I is the closest zone to the \textit{Active Carpet} and the fluid-fluid interface, where hydrodynamic fluctuations are dominantly vertical; Region II is further up from the \textit{Active Carpet} and is characterized by isotropic hydrodynamic fluctuations; Region III is the furthest region, near the free surface and is dominated by horizontal flow fluctuations. We demonstrate that the extent of these regions depends strongly on the degree of confinement, i.e. the layer thickness and the strength of the viscosity jump. Lastly, we show that confinement fosters the emergence of large-scale flow structures within the layer housing the \textit{Active Carpets}--not previously reported. Our findings shed light on the complex interplay between confinement and hydrodynamics in floating viscous film biological systems, providing valuable insights with implications spanning from ecological conservation to bio-inspired engineering.
\end{abstract}

\maketitle

\section{Introduction}

Microbial swimmers are ubiquitous across Earth's aquatic landscapes \cite{lauga2009hydrodynamics}. They thrive from wet surfaces to abyssal oceans, enduring extremes from freezing to boiling waters \cite{dechesne2010hydration,vetriani2014deep,doting2024exometabolome}. Microswimmers are finely equipped with an array of skills and taxes enabling them to navigate towards optimal regions or niches \cite[e.g.,][]{javadi2020photo,durham2012thin}, where they harness chemical compounds and sunlight to synthesize their own food, forming colonies known as `\textit{Active Carpets}' \cite{mathijssen2018nutrient,guzman2021active}. These ecological niches of optimal food uptake and growth are usually found near the interface between two layers. For instance, photosynthetic microbes habit around the thermocline, the zone of maximum vertical temperature gradient in the water column, separating the upper warmer layer from the deeper, nutrient-enriched colder layer \citep{sharples2001phytoplankton,durham2012thin,fernandez2021inhibited}. Microbial swimmers have adapted to the inherently layered nature of aquatic environments, where physical, chemical, and biological properties undergo a gradient in the direction of gravity. These layers can arise from various abrupt changes in fluid properties, including temperature (thermocline), salinity (haloclines), density (pycnoclines), dissolved oxygen (oxycline), dissolved chemicals (chemoclines), or sharp shifts in viscosity \cite{stocker2012marine, desai2020biofilms}. Moreover, most of Earth's aquatic systems are capped by the air-water interface, the free surface, where the atmosphere meets waters or thin oil films or slicks lying on top of water. Such a layering environment creates a wide range of niches for microbial swimmers and inert suspended matter \cite{ahmadzadegan2019hydrodynamic,deng2020motile,subbiahdoss2020biofilm,voskuhl2022natural}, as well as vertical confinement that may control the ability of microbes to harvest food and self-clean via collective hydrodynamic fluctuations.  
Biogenic hydrodynamic stirring has been proved to enhance active diffusion \cite{morozov2014enhanced,mino2013induced,sommer2017bacteria,mathijssen2017universal,sepulveda2021persistence,simoncelli2018biogenic,ran2021bacteria,singh2021bacterial}, generate persistent flows for feeding processes in collective and single microswimmers \cite{kanale2022spontaneous,mathijssen2016hydrodynamics,vskultety2023hydrodynamic}, produce aggregation \cite{belan2019pair,aguayo2023floating,gokhale2022dynamic,kushwaha2023phase,maheshwari2019colloidal,grossmann2024non} and induce long-range hydrodynamic fluctuations \cite{guzman2021active}. \textit{Active Carpets} formed by Stokeslets living close to non-slip boundaries drive non-equilibrium diffusion that drive an active diffusion much larger than thermal Brownian motion \cite{mathijssen2018nutrient,guzman2021active}. Moreover, the diffusion is tensorial, with a vertical diffusion larger than horizontal diffusion, and it produces non-equilibrium transport that does not follow the Boltzmann distribution. \citet{aguayo2023floating} recently studied the hydrodynamic fluctuations induced by \textit{Active Carpets} formed by dipole microswimmers organized into a thin layer beneath the free surface of a semi-infinite homogeneous fluid. The authors found that the hydrodynamics fluctuations produce anisotropic diffusion that decays slowly with distance, showing a remarkable biogenic long-range effect on vertical transport.

Vertical transport is essential for mixing and ventilation in stratified environments. A long-debated question is whether swimming organisms contribute to such processes in aquatic systems \cite{simoncelli2017can,kunze2019biologically}. To uncover the role of microorganisms in stratified liquid systems, \citet{ardekani2010stratlets} employed the multipole expansion technique to examine a single microswimmer, known as a stratlet. According to the biophysics operating at the micro-scale, the stratlet model considers an organism moving at low Reynolds numbers, $Re\ll 1$, in which viscous forces dominate over inertial ones. Their findings revealed that pullers moving in a stratified fluid do not generate persistent flows conducive to vertical mixing. Instead, they induce in-plane flows that strengthen the existing stratification. Theoretical studies further corroborated these findings, demonstrating that at low-$Re$, a single microbe swimming perpendicular to isopycnals has minimal impact on mixing \cite{wagner2014mixing}. Yet, numerical and laboratory experiments have demonstrated that the collective vertical migration of small organisms--like zooplankton swimming at intermediate Reynolds number $Re\sim O(1)$ \cite{wang2015biogenic,noto2023simple}--can drive a significant enhancement of the transport and mixing across fluid-fluid interfaces \cite{houghton2018vertically,ouillon2020active}. Moreover, recently, it has been theoretically predicted that \textit{Active Carpets} formed by large numbers of organisms swimming at low-$Re$ can drive long-range hydrodynamic fluctuations that enhance transport in homogenous fluid environments. This prompts us to delve into the behavior of hydrodynamic fluctuations driven by \textit{Active Carpets} within layered environments.

\begin{figure*}
\centering 
\includegraphics[width=1\textwidth]{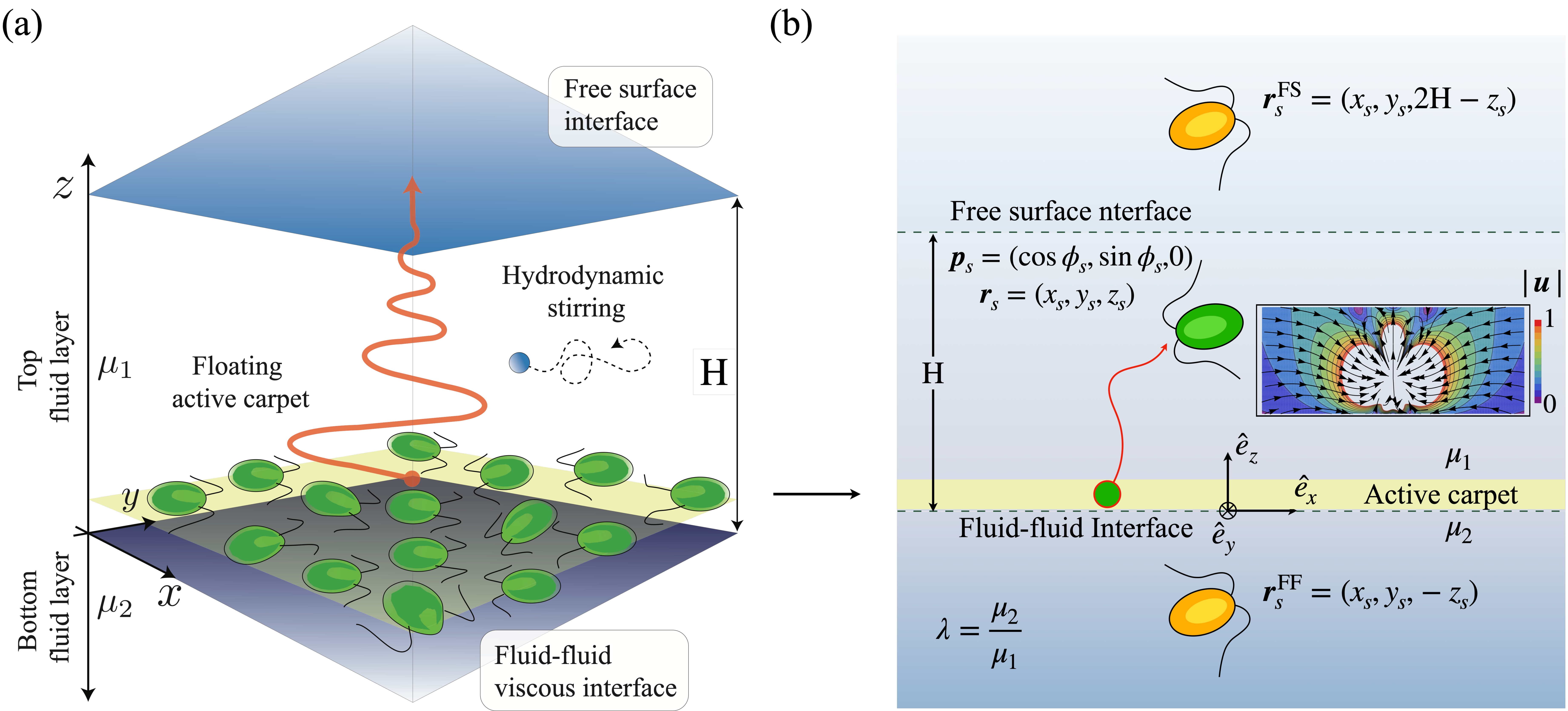}	
    \caption{
(a) Schematic of a confined \textit{Active Carpet} (yellow) formed by microswimmers living at a fixed distance $\sigma$ from the bottom fluid-fluid interface (in green). The microswimmers stir their surrounding confined fluid (orange arrows). The ratio between fluid viscosities is $\lambda = \mu_2/\mu_1$. (b) Conceptual model for a single microswimmer confined between the parallel free surface interface and the fluid-fluid interface. Microorganisms swim in the $x$-$y$ plane at $z=\sigma$. (c,d) Velocity field induced by a microorganism in the fluid of viscosity $\mu_{1}$.  The first is a top-view projection in the $x$-$y$ plane, and the second is a front-view in the $x$-$y$ plane.}
\label{Fig:Fig1}%
\end{figure*}

This manuscript focuses on \textit{Active Carpets} living within a confined layer, bounded above by a free surface (air-water interface) and below by a fluid-fluid interface characterized by a viscosity interface that separates two fluid layers of viscosities $\mu_1$ and $\mu_2$, as illustrated \Cref{Fig:Fig1}(a). Within this environment, an \textit{Active Carpet} experiences two confining mechanisms: geometrical confinement, characterized by the thickness of the upper layer $\rm H$ (the floating film), and viscous confinement, characterized by the viscosity jump or ratio $\lambda=\mu_2/\mu_1$. 
We can posit two scenarios for $\lambda$. In a first scenario, $\lambda<1$, the \textit{Active Carpet} lives in a fluid that it is more viscous that its underlying layer. Whereas in the second scenario, $\lambda>1$, the \textit{Active Carpet} lives in a less viscous layer than the subject layer. In this case, $\lambda$ tells us how soft $(\lambda\rightarrow 1)$ or rigid $(\lambda \rightarrow \infty)$ the fluid-fluid viscous interface is. 

Recent research has studied the microorganisms' optimal swimming behavior in viscosity gradients (viscotaxis), revealing divergent trends. In some cases, microorganisms prefer swimming in low-viscosity regions, while in others, they favor high-viscosity environments \cite{datt2019active,liebchen2018viscotaxis,shaik2021hydrodynamics, durham2012thin,marcos2012bacterial}. For instance, the \textit{Chlamydomonas reinhardtii}, a type of phytoplankton, shows viscophobic behavior by forming thin layers near interfaces with $\lambda<1$ \cite{stehnach2021viscophobic}.
Building upon the framework of \textit{Active Carpets} \cite{mathijssen2018nutrient,guzman2021active}, we explore theoretically and numerically the impact of confinement on hydrodynamic fluctuations resulting from the collective swimming of organisms forming an \textit{Active Carpet} residing at the upper vicinity of a viscosity interface. 

In \Cref{sec:2}, we introduce the mathematical model for both a single swimmer moving at low-$Re$ and an \textit{Active Carpet} in a layered environment. We explain the numerical modeling approach and parameters in \Cref{sec:3}. In \Cref{sec:4}, we present and discuss our main findings, comprising new analytical solutions for \textit{Active Carpets} in a layered aquatic environment, analyses of the topology of hydrodynamic fluctuations and the evidence of large-scale flow patterns driven by the \textit{Active Carpet} within the layered system. Our study aims to shed light on the complex interplay between confinement and hydrodynamics in these intriguing biological systems. 

\section{Mathematical model}\label{sec:2}
\subsection{A microswimmer in a layered environment}

We model the flow field of a single flagellated microswimmer moving in a homogeneous fluid of viscosity $\mu_1$, confined between two non-deforming boundaries \cite{aderogba1978action,desai2020biofilms}. Here, the top boundary is a free surface interface, whereas the bottom is a fluid-fluid interface, where below it there is a second fluid of viscosity $\mu_2$. The ratio between both fluid viscosities is denoted as $\lambda=\mu_2/\mu_1$. The distance between both interfaces is $\mathrm{H}$ and is referred to as the film thickness.

The low Reynolds number regime describes the fluid motion driven by microscopic swimmers. In this regime, viscous forces are larger than inertial ones \cite{taylor1951analysis,purcell1977life}. Here, hydrodynamics flows are governed by the incompressible Stokes equations \cite{stokes1851effect}. For a point force $\boldsymbol{F}=\boldsymbol{f}\delta (\boldsymbol{r}-\boldsymbol{r}_s)$ acting at $\boldsymbol{r}_s=(x_s,y_s,z_s)$ on a stationary fluid of dynamic viscosity $\mu$ we have 
\begin{subequations}\label{eq:stokes_eq}
    \begin{equation}
    \nabla p(\boldsymbol{r}) -\mu \nabla^2 \boldsymbol{u}(\boldsymbol{r}) =\boldsymbol{F}
    \end{equation}
    \begin{equation}
        \nabla\cdot\boldsymbol{u}(\boldsymbol{r})=0
    \end{equation}
\end{subequations}
where $\boldsymbol{u}(\boldsymbol{r})$ and $p(\boldsymbol{r})$ are the fluid velocity and pressure fields at position $\boldsymbol{r}=(x,y,z)$, respectively. The solution of \Cref{eq:stokes_eq} in the absence of boundaries is  well-known as the Stokeslet \cite{hancock1953self,happel2012low,kim2013microhydrodynamics}

\begin{equation}
\boldsymbol{u}(\boldsymbol{r},\boldsymbol{r}_s)=\frac{\mathcal{G}_{ij}(\boldsymbol{r},\boldsymbol{r}_s)\cdot \boldsymbol{F}}{8\pi\mu},\;\mathcal{G}_{ij}(\boldsymbol{r},\boldsymbol{r}_s)
   = \left( \dfrac{\delta_{ij}}{|\boldsymbol{d}|}+\dfrac{d_i d_j}{|\boldsymbol{d}|^3} \right),
\end{equation}
where $\boldsymbol{d} = \boldsymbol{r}-\boldsymbol{r}_s$ is the relative position of the swimmer to the fluid.
Derivatives of the Stokeslet are also solutions \cite{mathijssen2015tracer,spagnolie2012hydrodynamics,chwang1975hydromechanics}. For swimming microorganisms, where the thrust and the drag force balance, the Stresslet is the first dominant term in the multipole expansion to describe their far-field flow \cite{batchelor1970stress}. 

Next, we include the boundaries. Following \citet{desai2020biofilms}, we construct the first approximation to the image system for the flow field produced by a microswimmer confined between a fluid-fluid interface and a free surface interface, as shown in \Cref{Fig:Fig1}(b).

The Stokes equations for a point force acting on fluid 1 are
\begin{subequations}\label{eq:Stokes_eq_1}
    \begin{equation}
        -\nabla p^{(1)} + \mu_1 \nabla^2 \boldsymbol{u}^{(1)} + \boldsymbol{F}=0,
    \end{equation}
    \begin{equation}
        \nabla \cdot \boldsymbol{u}^{(1)} = 0,
    \end{equation}
\end{subequations}
while for fluid 2 are
\begin{subequations}\label{eq:Stokes_eq_2}
    \begin{equation}
        -\nabla p^{(2)} + \mu_2 \nabla^2 \boldsymbol{u}^{(2)}=0,
    \end{equation}
    \begin{equation}
        \nabla \cdot \boldsymbol{u}^{(2)} = 0.
    \end{equation}
\end{subequations}

The velocity field of both fluids must satisfy the following boundary conditions at the fluid-fluid interface \cite{aderogba1978action},
\begin{subequations}\label{ffeqs}
    \begin{equation}
        {u_\alpha}^{(1)} =  {u_\alpha}^{(2)},\, \text{at} \, z=0
    \end{equation}
    \begin{equation}
        u_z^{(1)} = u_z^{(2)} =0,
    \end{equation}
    \begin{equation}
        \mu_1 \left(\dfrac{\partial u_{\alpha}^{(1)}}{\partial z}+\dfrac{\partial u_{z}^{(1)}}{\partial \alpha}\right) = \mu_2 \left(\dfrac{\partial u_{\alpha}^{(2)}}{\partial z}+\dfrac{\partial u_{z}^{(2)}}{\partial \alpha}\right) \, \text{at} \, z=0,
    \end{equation}
\end{subequations}
where $\alpha=x,y$, while at the free surface interface,
\begin{subequations}
    \begin{equation}
        u_z^{(1)}= 0,\, \text{at}\, z=\mathrm{H},
    \end{equation}
    \begin{equation}
        \dfrac{\partial u_{\alpha}^{(1)}}{\partial z}+\dfrac{\partial u_{z}^{(1)}}{\partial \alpha}=0,\, \text{at} \,z=\mathrm{H}.
    \end{equation}
\end{subequations}

From solving \cref{eq:Stokes_eq_1,eq:Stokes_eq_2,ffeqs}, the image's system for an interface between two fluids of different viscosities leads to a generalization of the Blake tensor \cite{aderogba1978action}
\begin{equation}
\begin{split}
    \mathcal{A}_{ij}(\boldsymbol{r}) &= \left(\dfrac{1-\lambda}{1+\lambda} \delta_{j\alpha} \delta_{\alpha k} -\delta_{j3} \delta_{3k}\right)\mathcal{G}_{ik}(\boldsymbol{r}) \\
    &+ \dfrac{2\lambda}{\lambda+1}h (\delta_{j\alpha}\delta_{\alpha k}-\delta_{j3}\delta_{3k})\dfrac{\partial}{\partial r_k} \left(\dfrac{h r_i}{r^3}+ \mathcal{G}_{i3}(\boldsymbol{r})  \right).
    \label{eq:tensorA}
\end{split}
\end{equation}
For $\lambda= 0$, we recover the image flow field of a point force close to a free surface,
\begin{equation}
    \mathcal{F}_{ij}(\boldsymbol{r}) = \mathcal{M}_{jk}\mathcal{G}_{ik}(\boldsymbol{r}),
    \label{eq:mirror}
\end{equation}
where $\mathcal{M}_{jk}$ is a mirror matrix, $\mathcal{M} = \text{diag}(1,1,-1)$. This image system corresponds to the one satisfying the boundary condition at $z=\mathrm{H}$.

Formally, a point force $\boldsymbol{F}=\boldsymbol{f}\delta(\boldsymbol{r}-\boldsymbol{r}_s)$ is acting at position $\boldsymbol{r}_s=(x_s,y_s,z_s)$. The image system to account for the fluid-fluid interface will be the one from \cref{eq:tensorA}, located at position $\boldsymbol{r}_s^{\mathrm{FF}} = (x_s,y_s,-z_s)$. On the other hand, to account for the free surface interface we set a mirror image from \cref{eq:mirror} at position $\boldsymbol{r}_s^{\mathrm{FS}}=(x_s,y_s,2\mathrm{H}-z_s)$. The velocity field produced by a point force parallel to both interfaces is given by 
\begin{equation*}
    \begin{split}
     \boldsymbol{u}(\boldsymbol{r},\boldsymbol{r}_s,\lambda,\mathrm{H}) =& \mathcal{G}(\boldsymbol{r},\boldsymbol{r}_s)\cdot\boldsymbol{f_{\parallel}}\\ 
     &+ ( \mathcal{F}(\boldsymbol{r},\boldsymbol{r}_s^{\mathrm{FS}},\mathrm{H})+\mathcal{A}(\boldsymbol{r},\boldsymbol{r}_s^{\mathrm{FF}},\lambda) )\cdot\boldsymbol{f_{\parallel}},
    \end{split}
\end{equation*}
where the second term is the full image system added to satisfy the induced hydrodynamic boundary conditions and $\boldsymbol{f_{\parallel}}= \boldsymbol{f}/(8\pi \mu_1)$ is the scaled force. 

As shown in \Cref{Fig:Fig1}(b), because of the linearity of the Stokes equations, we use the same image system for expressing the velocity field of a force dipole with orientation $\boldsymbol{p}_s$,
\begin{equation}
\boldsymbol{u}_D(\boldsymbol{r},\boldsymbol{r}_s,\boldsymbol{p}_s) = \kappa (\boldsymbol{p}_s\cdot \nabla_s)[(\mathcal{G}+\mathcal{F}+\mathcal{A})\cdot\boldsymbol{p}_s],
    \label{dipolepuller}
\end{equation}
where derivatives $\nabla_s = \partial/\partial \boldsymbol{r}_s$ are taken with respect to the microswimmer's position.
Here, $\kappa=f_{\parallel}\sigma/\mu_1$ corresponds to the dipolar strength, which characterizes the hydrodynamic distortions of each microswimmer and sets the timescale of the biogenically driven flow, $\tau=\sigma^3/\kappa$, with $\sigma$ the microswimmer body length. On the one hand when $\kappa>0$, the microswimmer generates an extensile flow (pusher) similar to motile bacteria powered by a helical flagella bundle. On the other hand when $\kappa<0$, the flow field is contractile (puller) similar to the flow generated by green microalgae {\it Chlamydomonas reinhardtii}, which forms thin phytoplankton layers in aquatic environments \cite{durham2012thin}. 

In nature, sharp vertical temperature gradients or the presence of natural and artificial oils can lead to sharp contrasts in fluid viscosity; especially at the skin of surface waters \cite{hondzo2022measurement}. Also, during blooms, planktonic microorganisms have been reported to increase the effective ambient viscosity, generating strong viscosity gradients with relative viscosities $1\lesssim\lambda\lesssim 3$ at short distances \cite{guadayol2021microrheology}. On the contrary, bacterial suspension has been proved to decrease ambient viscosity, depending on their concentration \cite{martinez2020combined}, with relatives viscosities $O\left(10^{-3}\right)\lesssim \lambda \lesssim O(1)$.

In this study, we utilize the leading-order image system to describe far-field flows \cite{desai2020biofilms}. It is important to note that when two image flow fields are added to the hydrodynamic system, boundary conditions are only partially satisfied. To mitigate this discrepancy, recursive images of the original images must be added. However, since the first images closely approximate the confined natural system under investigation, they encapsulate most of the relevant physics. Subsequent images will inevitably be farther removed from the system \cite{mathijssen2016hydrodynamics}.

As an example, the flow field generated, in the $x$-$z$ plane, by a single puller microswimmer in a film of thickness $\mathrm{H=3}$, with $\kappa=30$, the dipole strength of a flagellated microswimmer moving in water reported by Drescher~et~al.~\cite{drescher2010direct}, and viscosity ratio $\lambda=1.5$ -- computed from \cref{dipolepuller} -- is shown in the inset of \cref{Fig:Fig1}(b). Streamlines show the typical trajectories followed by fluid parcels around this type of swimmer close to surfaces \cite{drescher2010direct,mathijssen2016hydrodynamics}; contours show the intensity of the velocity field, this being more intense close to the singularity (white) and weaker far from it (purple).

\subsection{\textit{Active Carpet} in a layered aquatic environment}
\label{sec:Active Carpet in a layered aquatic environment}

Our model considers a collection of dipole microswimmers constricted to move in the $x$-$y$ plane at a fixed height $z=\sigma$, which corresponds to microswimmers moving above the viscosity interface forming the so-called Active Carpet, as sketched in \cref{Fig:Fig1}(a). The \textit{Active Carpet} is constituted of a dense suspension of microorganisms, each of them inhabiting this environment at positions ${\bm r}_s=(x_s,y_s,\sigma)$ and orientations ${\bm p}_s=(p_x,p_y,0)$. We consider them in a stroboscopic diffusive regime \cite{aguayo2023floating, guzman2021active} such that as time progresses, all microswimmers are uniformly distributed in space and orientations in a 2D surface. Each microorganism stirs and energizes the confined water film, driving a flow field given by \cref{dipolepuller}. Here, we probe the biogenically driven flow at positions ${\bm r}_0=(x_0,y_0,z_0)$, corresponding to a fluid parcel between the free surface and fluid-fluid interface. To find a far-field approximation for probe fluid parcels far from the Active Carpet, we performed a Taylor expansion for the flow field generated by a single microswimmer, such that $z_s =\epsilon z$ with $\epsilon\ll 1$. With this analytical expression, we can measure different statistical properties of the collectively generated flow.

The mean flow field averaged over a finite carpet of size $R$ with microswimmer uniformly distributed in cylindrical coordinates, $\rho_s, \;\theta_s$ and orientations $\phi_s$ is
\begin{equation}
\label{eq:flow}
    \langle \boldsymbol{v}(\boldsymbol{r}) \rangle = \int \boldsymbol{u} (\boldsymbol{r},\boldsymbol{r}_s,\boldsymbol{p_s})\, \mathcal{F} (\boldsymbol{r}_s,\boldsymbol{p_s})\, d \boldsymbol{r}_s d \boldsymbol{p_s},
\end{equation}
where $\mathcal{F} = n/2\pi>0$ is a uniform distribution of swimmers for a carpet number density $n$ and $\langle \cdot\rangle$ is the average  over the collection of swimmers. Here $\boldsymbol{r_s}=(\rho_s \cos{\theta_s},\rho_s \sin{\theta_s},\sigma )$ is the swimmer's position and $\boldsymbol{p_s} = (\cos{\phi_s},\sin{\phi_s},0)$ the swimmer's orientation, with $\phi_s,\theta_s \in [-\pi,\pi]$. Such a carpet is able to generate a mean flow field that attracts suspended particles \cite{mathijssen2018nutrient}. However, when the carpet size increases significantly, i.e. $R\rightarrow \infty$, the mean flow converges to zero \cite{guzman2021active}. Here, we consider a infinite carpet, so that $ \langle \boldsymbol{v}(\boldsymbol{r}) \rangle=0$. Yet, the variance $ \mathcal{V}_{ij}= \langle \boldsymbol{v}_i \boldsymbol{v}_j\rangle$ is different from zero, where $i,j=x,y,z$ denote its components in cartesian coordinates.
The variance, which encapsulates what we call fluctuations, governs the active diffusion process that tracer particles experience over the carpet; its magnitude depends strongly on the geometry of the environment and can be computed analytically using the following expression for the variance tensor 
\begin{equation}
   \mathcal{V}_{ij} =  \langle v_i v_j \rangle = \int u_i u_j \mathcal{F}\, d \boldsymbol{r}_s d \boldsymbol{p_s}.
   \label{eq:variance}
\end{equation}

In addition, we define the average vorticity field induced by the confined \textit{Active Carpet} as
\begin{equation}
     \langle \boldsymbol{\omega} \rangle=\langle \nabla \times \boldsymbol{v} \rangle = \int (\nabla \times \boldsymbol{u}) \mathcal{F}\, d \boldsymbol{r}_s d \boldsymbol{p_s}.
   \label{eq:vorticity}
\end{equation}

We now turn the focus to the numerical framework utilized to simulate the dynamics of confined \textit{Active Carpet}.

\section{Numerical simulations}
\label{sec:3}
As discussed in \Cref{sec:Active Carpet in a layered aquatic environment}, we do not follow each microswimmer's trajectory in time. Instead, we investigate the effects of the \textit{Active Carpet} in the surrounding fluid once microswimmers have passed from the ballistic regime to a diffusive regime \cite[see e.g.][]{guzman2021active,aguayo2023floating}.
In brief, we model \textit{Active Carpets} through the Fast Dynamics Framework. The Fast Dynamics Framework is made by randomly distributing $N_{s}=10^5$ microswimmers with positions $\boldsymbol{r}_{s}=(x_i,y_i,\sigma)$ and orientations $\boldsymbol{p}_s = (\cos \phi_i,\sin \phi_i,0)$, with $\phi_i \in [-\pi,\pi]$, $i \in [1,N_s]$ in a finite square domain, $L \times L $, where $L$ is the size of the \textit{Active Carpet}. The microswimmers number density in the \textit{Active Carpet} is defined as $n = N_{s} /(2L)^2$. 
Both the positions and orientations are uniformly distributed. 
The total flow generated by the active carpet is computed by superposing the flow velocity of each individual microswimmer given by \cref{dipolepuller}. Thus, the collective velocity field driven by the ensemble, or the colony, is determined by
\begin{equation}
    \boldsymbol{v}(\boldsymbol{r},\boldsymbol{r}_s,\boldsymbol{p}_s) = \sum_{i=1}^{N_{s}} \boldsymbol{u}_D(\boldsymbol{r},\boldsymbol{r}_s,\boldsymbol{p}_s).
    \label{eq:total_v}
\end{equation}
This velocity is then computed for an ensemble of $N_e$ independent \textit{Active Carpet} configuations.
Hence, we compute the ensemble variance tensor of the flow field \cref{eq:total_v}, as defined in \cref{eq:variance}, averaging over a finite number of \textit{Active Carpet} ensembles $N_{e}$. Using this framework, we investigate the dynamics of passive tracer particles stirred by the collective action of the colony. A tracer particle, with position $\boldsymbol{r}^T(\tau_e)$, feels the aggregated flow exerted by the \textit{Active Carpet} ensemble and reacts to changes in the re-orientation microswimmers experience as time progresses;  $\tau_e$ is the dimensionless time. As a result, passive tracer dynamics are governed by the following equation \cite{pushkin2013fluid,mathijssen2015tracer,guzman2021active,aguayo2023floating}
\begin{equation}\label{eq:tracerdyn}
    \dfrac{d\boldsymbol{r}^{T}(\tau_e)}{d \tau_e} = \boldsymbol{v}(\boldsymbol{r}^{T},\boldsymbol{r}_s,\boldsymbol{p}_s).
\end{equation}
As illustrated in \Cref{Fig:Fig1}(a), we explore tracer particle dynamics between the \textit{Active Carpet} and the surface interface, $\sigma<\boldsymbol{r}^{T}(\tau_e)<\mathrm{H}$. We integrate \cref{eq:tracerdyn} numerically utilizing an Euler scheme, with a nondimensional integration time step $\Delta t = 10^{-3}$. At each time step, the tracer particle excursions a length $\Delta \boldsymbol{r}^T$ determined by a new random independent \textit{Active Carpet} ensemble.

The size of the \textit{Active Carpet} is crucial in computing every observable within this framework. \citet{mathijssen2018nutrient} stressed that if the \textit{Active Carpet} is too small, a drift flow can overpower the motion of tracer particles. To mitigate this effect, we ensure that the \textit{Active Carpet} is sufficiently long so that tracer trajectories are controlled by the biogenically driven fluid flow \cite{mathijssen2018nutrient,guzman2021active,aguayo2023floating}.

In this study, we set the following parameters unless otherwise stated: dipole strength $\kappa=-30$ following \cite{drescher2011fluid}, \textit{Active Carpet} number density $n=0.1$, vertical position in the far-field approximation $\sigma=1$, small parameter $\epsilon=0.1$, ensemble number $N_e=10^3$, and \textit{Active Carpet} length $L=10^4$. Our primary focus is investigating the impact of confinement size $\mathrm{H}$ and interface strength $\lambda$. In what follows, we explore how the degree of fluid confinement impacts the spatial characteristics of the hydrodynamic stirring driven by \textit{Active Carpets}.

\section{Results and Discussion}\label{sec:4}

\subsection{Fluctuation and dispersion of tracer particles}\label{sec:4a}
We first characterize the hydrodynamic fluctuations driven by the confined \textit{Active Carpet} whose members swim parallel to the boundary interfaces, as sketched in \Cref{Fig:Fig1}(a). To this point, the theory developed by \citet{mathijssen2018nutrient} allows us to directly compute the variance \cref{eq:variance} of the flow field \cref{dipolepuller} by performing a far-field approximation of the flow field (\Cref{sec:Active Carpet in a layered aquatic environment}). Note that off-diagonal components of the variance tensor vanish. Additionally, by symmetry, we have that  $\langle v_{x}^2\rangle=\langle v_{y}^2\rangle$ \cite{guzman2021active}. Therefore we do have only two fluctuations to characterize, the horizontal $\langle v_{x}\rangle$ and the vertical one, $\langle v_{z}^2\rangle$. 

As a consequence of the confined geometry and the boundary conditions, the velocity field and its variance depend upon the ratio $\lambda=\mu_{2}/\mu_{1}$, characterizing the viscous interface, the thickness of the aquatic film $\mathrm{H}$, as well as the intrinsic physical and geometrical parameters of the Stokes solution and far-field approximation. Hence, each variance component is a function of $\langle v_{i}^2\rangle = \langle v_{i}^2\rangle(n,\kappa,\sigma,\lambda,\epsilon,\mathrm{H},z_0)$, where $z_0$ is the height of a fluid parcel relative to the \textit{Active Carpet}. The obtained analytical expressions are
\begin{widetext}
\begin{subequations}\label{eq:vxyz}
\begin{equation}\label{eq:vxy}
\begin{split}
 \langle v_{x}^2\rangle & = \dfrac{\pi \kappa^2 n}{64} \left(\dfrac{21 \sigma^2 \epsilon^2-22 \sigma \epsilon  \left(z_0-2 \mathrm{H}\right)+11 \left(z_0-2
   \mathrm{H}\right)^2}{\left(z_0-2 \mathrm{H}\right)^4}+\dfrac{336 \sigma^2 \lambda ^2 \epsilon ^2}{(\lambda +1)^2 z_0^4}+\dfrac{160 \sigma
   \lambda  \epsilon }{(\lambda +1)^2 z_0^3} \right.\\
 &+\frac{36 z_0 (8 \sigma \lambda  \epsilon  (\mathrm{H}+2 \sigma \epsilon )+\mathrm{H} (2 \mathrm{H}+3 \sigma \epsilon
   ))}{\mathrm{H}^5 (\lambda +1)}-\dfrac{36 z_0^2 (2 \sigma \lambda  \epsilon  (2 \mathrm{H}+5 \sigma \epsilon )+\mathrm{H} (\mathrm{H}+2 \sigma \epsilon
   ))}{\mathrm{H}^6 (\lambda +1)}\\
   &\left.+\frac{8 \mathrm{H} (\mathrm{H}+\sigma \epsilon )-32 \sigma \lambda  \epsilon  (2 \mathrm{H}+3 \sigma \epsilon )}{\mathrm{H}^4
   (\lambda +1)}+\dfrac{44}{(\lambda +1)^2 z_0^2}\right),
   \end{split}
\end{equation}
\begin{equation}\label{eq:vz}
\begin{split}
    \langle v_{z}^2\rangle & =  \dfrac{9 \pi \kappa^2 n}{32}\left( \dfrac{2 \sigma^2 \epsilon ^2-2 \sigma \epsilon  \left(z_0-2 \mathrm{H}\right)+\left(z_0-2
   \mathrm{H}\right)^2}{\left(z_0-2 \mathrm{H}\right)^4}+
    \dfrac{80 \sigma^2 \lambda ^2 \epsilon ^2}{(\lambda +1)^2 z_0^4}+\dfrac{32 \sigma \lambda
    \epsilon }{(\lambda +1)^2 z_0^3} \right.\\
   & -\dfrac{4 z_0 (8 \sigma \lambda  \epsilon  (\mathrm{H}+2 \sigma \epsilon )+\mathrm{H} (2 \mathrm{H}+3 \sigma \epsilon
   ))}{\mathrm{H}^5 (\lambda +1)}+\dfrac{4 z_0^2 (2 \sigma \lambda  \epsilon  (2 \mathrm{H}+5 \sigma \epsilon )+\mathrm{H} (\mathrm{H}+2 \sigma \epsilon
   ))}{\mathrm{H}^6 (\lambda +1)} \\
   &\left.  +\dfrac{4}{(\lambda +1)^2 z_0^2} 
    \right).
    \end{split}
\end{equation}
\end{subequations}
\end{widetext}
Note that $\epsilon\ll 1$ but not zero. The limit $\epsilon\rightarrow 0$ does not have a physical interpretation. To validate these analytical results we performed simulations using \cref{eq:total_v} to obtain numerically the total velocity field averaged over $N_e$ of \textit{Active Carpet} ensembles (\Cref{sec:3}). From numerical results, we computed the variance for a set of $z_0$ values within the range $(\sigma,\mathrm{H})$, where we set the body length $\sigma$ as the unit of length through our results. 
\begin{figure*}[ht]
    \centering 
\includegraphics[width=\textwidth]{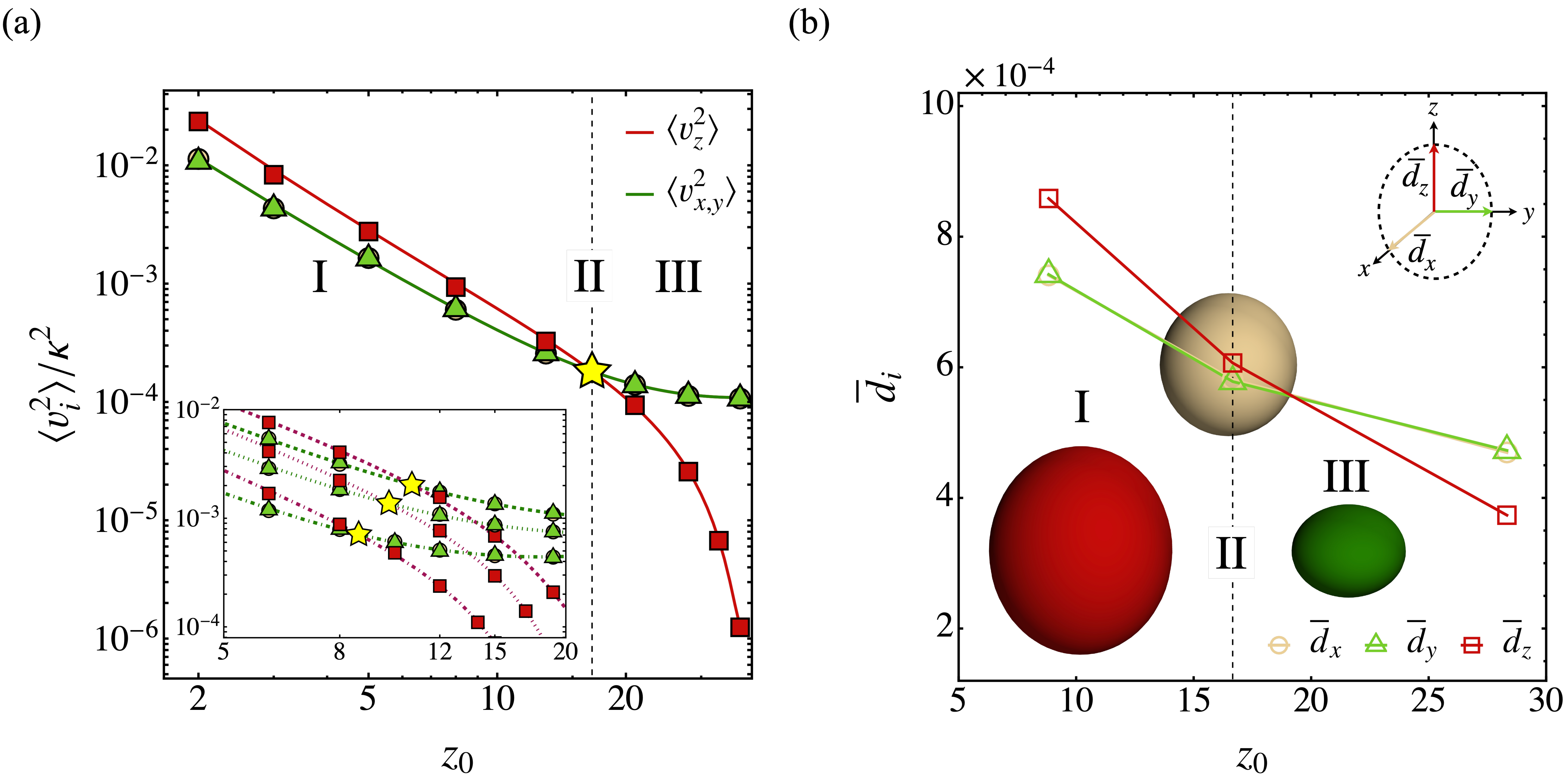}	
    \caption{
    Anisotropic diffusion driven by \textit{Active Carpets}. 
    (a) Variance of the hydrodynamic fluctuations driven by an \textit{Active Carpet} in the horizontal and vertical directions (green, red solid lines), as defined in \Cref{eq:variance}. Markers are simulation points obtained from \cref{eq:total_v}. The star marks the intersection between theoretical variances $\langle v_z^2\rangle(z_{\star})=\langle v_{xy}^2\rangle(z_{\star})$. Regions $\rm I,II$ and $\rm III$ indicate the dominance of fluctuations according to the distance from the Active Carpet. Here $\mathrm{H}=40$ and $\lambda=1.5$. Inset: Theoretical variances for $\mathrm{H}=20$, and three different values of $\lambda$, $\lambda=0.1,0.5,1.5$, denoted by dashed, dotted and dash-dotted lines, respectively. Stars are cross-points. (b) Ellipsoids represent the average displacement of tracer particles computed from \cref{eq:tracerdyn} for each fluctuation region. In green, tracer particles start at $z_0<z_\star$, while in red, they start at $z_0<z_\star<\mathrm{H}$. In ginger, they start exactly at $z_0=z_\star$. }
\label{Fig:Fig2}%
\end{figure*}

\Cref{Fig:Fig2}(a) presents the analytical and simulation results for $\lambda=1.5$ and $\mathrm{H}=40$. First, the theory and simulations show great agreement. Second, the effect of confinement in the biogenically driven fluid flow is severe. As the plot shows, horizontal and vertical fluctuations cross at a particular height, which we denote as $z_0=z_\star$. We found that vertical variances are more significant than horizontal ones for almost half the film thickness, $z_0 \propto 0.43 \mathrm{H}$. Using Eqs. \Cref{eq:vxy,eq:vz} we can determine exactly that they cross at height $z_{\star}=16.6$ (represented by the yellow star on the plot); here, we expect the fluctuations to have the same strength. To achieve this, horizontal fluctuations become more significant than for heights closer to the Active Carpet, $z_{0}<z_{\star}$. According to this unique feature, we can separate the space into three regions, denoted by \textrm{I}, \textrm{II}, and \textrm{III}. Region \textrm{I} represents the range of heights where vertical fluctuations dominate over horizontal fluctuations, Region \textrm{II} represents the locus where fluctuations are isotropic, whereas Region \textrm{III} represents the range of heights where horizontal fluctuations dominate over vertical fluctuations. To illustrate this feature, the inset shows results of fluctuations for different values of $\lambda$ in the list [0.1,0.5,1.5] (dashes, dotted and dot-dash, respectively) and a film thickness $\mathrm{H}=20$. The variance curves intersect in the same fashion. Results show that the cross-point $z_{\star}$ varies depending on the value of $\lambda$. Smaller values of $\lambda$ lead cross-points closer to the top free surface, and the contrary happens for greater values, i.e. the cross-point gets closer to the Active Carpet.

Our results contrast significantly with previous studies of \textit{Active Carpets}. In the case of \citet{guzman2021active}, in which the authors investigated \textit{Active Carpets} living near a no-slip surface, the fluctuations decay monotonically as $\langle v_i^2 \rangle \propto z_0^{-4}$ for parallel dipoles. Similarly, \citet{aguayo2023floating} found that fluctuations were long-ranged,  with $\langle v_i^2 \rangle \propto z_0^{-2}$ in the case of a free surface. In our case, the hydrodynamic fluctuations decay in a more complex manner as a result of the confinement and the influence of the viscous fluid-fluid interface. Although all the variances maintain their anisotropic behavior, we observe a non-monotonic trend, with a striking change in the amplitude of the variance components and the dominance of directional effects. Indeed, at some distance $z_{\star}$, the vertical fluctuations decay much more rapidly with $z_{0}$ than the horizontal fluctuations as we go farther from the \textit{Active Carpet}. Therefore, characterizing the locus $z_\star$ and its dependence on the parameters controlling the degree of confinement is essential since it determines where fluctuations and the flow direction become spatially biased. 

We carried out simulations using \cref{eq:tracerdyn} to prove that the anisotropic behavior of fluctuations effectively impacts surrounding tracer particles depending on the region where they move. For this, we created an \textit{Active Carpet} with $N_s=10^5$ microswimmers and $L=1.5\times10^4$. The \textit{Active Carpet} was confined within a film of thickness $\mathrm{H}=40$, and the bottom interface was characterized by a viscosity ratio to $\lambda=1.5$. From the analytical solution \cref{eq:vxyz}, we determined the intersection height $z_\star$ by resolving the equation $\langle v_x^2 \rangle(z_{\star})=\langle v_z^2 \rangle(z_{\star})$. Once determined $z_{\star}$, we seeded passive tracers at a height $z_0 = 27.7 > z_\star$ to measure vertical-dominated motions, at a height $z_{0}=z_{\star}$ to measure isotropic motions, and a height $z_0 = 10 <z_\star$ to measure horizontal-dominated motion. Then, we simulated $\tau_{e} = 180$ time steps by numerically integrating \cref{eq:tracerdyn}. This time was enough to determine the characteristic topology of the space through which tracers are transported by the hydrodynamic fluctuations. The latter was obtained by averaging the displacement $\overline{d}_i$ for all tracer particles in every direction $\hat{e}_i$, $i=x,y,z$. The ellipsoids in \Cref{Fig:Fig2}(b) (scaled for visualization) show the resulting topology for each region. Thus, we confirm that tracer particles have a more prominent vertical motion at heights $z_0 < z_\star$ (red) closer to the Active Carpet, equal to horizontal motions at $z_0 = z_\star$ (ginger) and that horizontal motions become dominant at $z_0 > z_\star$ (green). This particular example demonstrating a zonification of the hydrodynamic fluctuations across the fluid layer raises more general questions. How does the degree of confinement, determined by $\rm H$ and $\lambda$, impact the geometry of the biogenically driven hydrodynamic fluctuations? And how does $z_{\star}$ change in terms of $\lambda$ and $\rm H$?

\begin{figure*}
    \centering 
\includegraphics[width=1\textwidth]{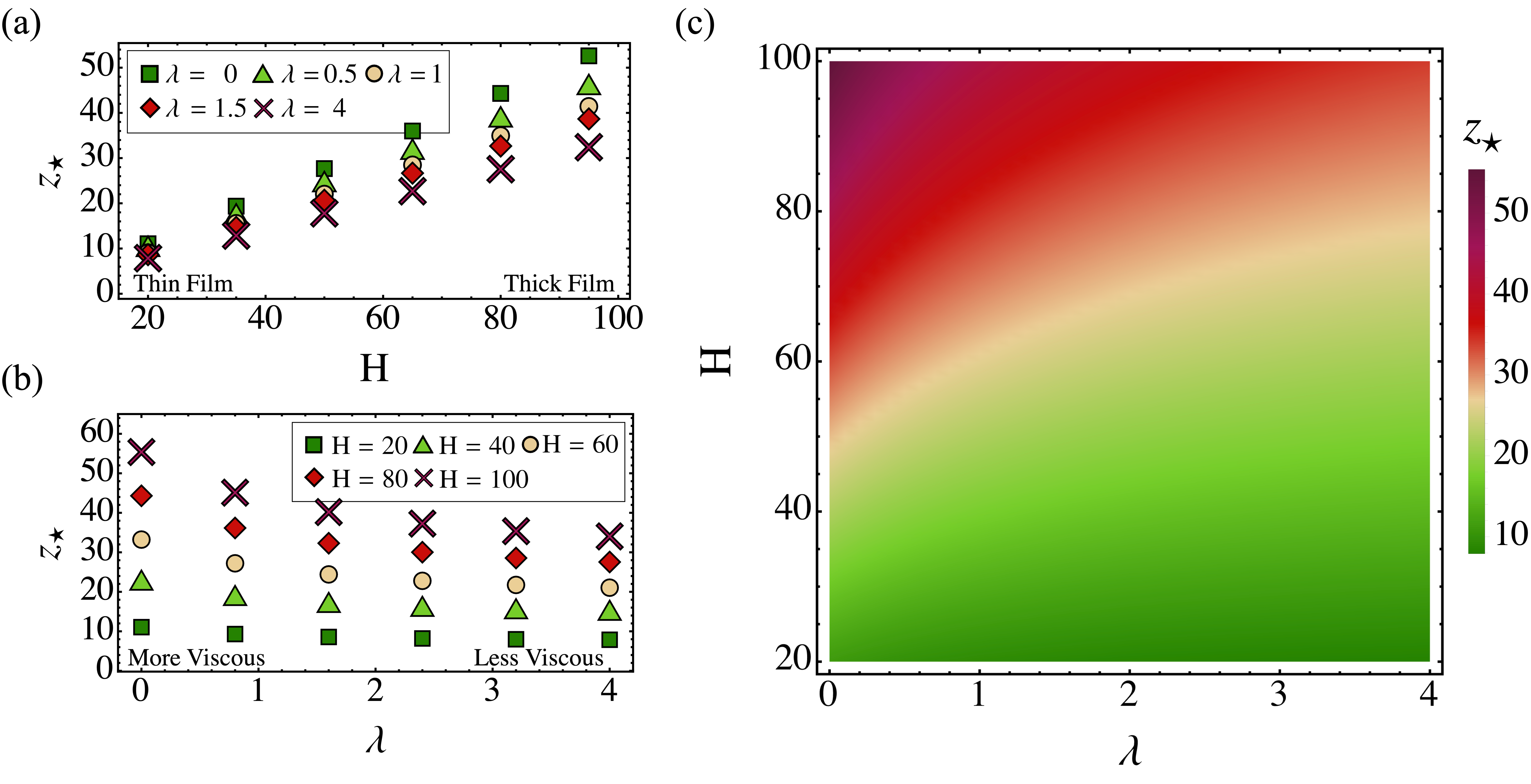}	
    \caption{
    The impact of confinement of the flow structure driven by \textit{Active Carpets}.
    Shown are analytical solutions for the cross-point height $z_\star$ for a domain of the parameters $\lambda$ and $\mathrm{H}$. In (a) from thin (small $\mathrm{H}$)  to thick films (large $\mathrm{H}$), and in (b) more viscous floating film ($\lambda <1$) to less viscous floating film ($\lambda >1$). In (c), the full dependence is shown.}
\label{Fig:Fig3}
\end{figure*}

\subsection{Impact of confinement on fluctuation's geometry}\label{sec:4b}

We explore answers to the above questions by examining how $z_\star$ depends upon the film's thickness $\mathrm{H}$ and the viscous fluid-fluid interface, $\lambda$. \Cref{Fig:Fig2} shows us a significant anisotropy and non-monotonic behavior of the hydrodynamic fluctuations induced by the confined Active Carpet. A major result is that we found the existence of three regions whose spatial distribution is determined by $z_\star$; the height above the \textit{Active Carpet} at which the vertical and horizontal variance of the velocity fluctuations equal, heights $z$ lower than $z_{\star}$ where vertical fluctuations dominate over horizontal fluctuations, and heights between $z_{\star}$ and $\rm H$, horizontal fluctuations dominate over vertical fluctuations. The analytical solution \cref{eq:vxyz} reported in \Cref{sec:4a} enables us to describe and analysis these three regions change as function of $\rm H$ and $\lambda$.

We computed $z_{\star}(\lambda,\mathrm{H})$
for a wide range of film's thicknesses, varying $H$ two order of magnitude, and varying the viscous fluid-fluid interface ratio $\lambda$ over a range that could potentially be observed in aquatic environments. For this, we resolve numerically the nonlinear equation for $\langle v^{2}_{x}\rangle(z,{\rm H},\lambda)-\langle v^{2}_{z}\rangle(z,{\rm H},\lambda)=0$ for $z\in(\sigma,H)$. The root of this equation corresponds to $z_{\star}$. 

\Cref{Fig:Fig3}(a) shows the dependency of $z_\star$ on the film's thickness $\mathrm{H}$ for five values of $\lambda\in[0,4]$. First, we identify that $z_{\star}$ increases monotonically with $10\leq H\leq 100$. This implies that thicker aquatic films create conditions for isotropic fluctuations to be further away from the Active Carpet, i.e. the region where vertical fluctuations dominate over horizontal fluctuations also expands monotonically with $H$. From an ecological viewpoint, this is relevant since the \textit{Active Carpet} can have longer range for both attracting or repealing mass. Second, we observe that $z_{\star}$ is larger for smaller $\lambda$. In other words, the more viscous is the layer where the \textit{Active Carpet} habits, the longer is the extent of the region where vertical fluctuations dominate over horizontal fluctuations. For the extreme case of $\lambda\rightarrow 0$, the fluid where the \textit{Active Carpet} inhabits is dramatically more viscous than the fluid below the \textit{Active Carpet}, $\mu_1 \gg \mu_2$. 

Such a big difference in viscosity can be observed in fresh and marine aquatic environments, where algae blooms change the physico-chemical properties of a region in water column \cite{mrokowska2022effect}. At some point, this dramatic spatial changes in viscosity push the fluid away from the Newtonian limit. Nonetheless, Newtonian rheology remains a fair approximation for gently varying fluid viscosity, as the case of liquid water \cite{mazza2016physics}. For $\lambda =0$, and the range of $\rm H$ values here explore, we obtain that $z_\star= 0.55 \mathrm{H}$. Whereas for a slightly larger value, $\lambda=0.5$, we obtain that $z_\star= \mathrm{H}/2$. We show that with low $\lambda$ values, the system is at a sort of local equilibrium phase of the hydrodynamics fluctuations effects. This means that fluctuations regions I and III will have almost of the same extent. At larger values of $\lambda$, the proportionality we just described changes. \Cref{Fig:Fig3}(a) shows a different result for $\lambda=4$, where we find that in average, $z_\star = 0.35 \mathrm{H}$. Here, we are strengthening the interface by making $\lambda$ larger; from our framework, one would expect to recover a well-studied hydrodynamic system, a free-surface liquid film held by a rigid surface for $\lambda\gg 1$ \cite{mathijssen2016hydrodynamics}. As a result, we observe that for larger $\lambda$ the hydrodynamic fluctuations are, on average, more planar-directed motions than vertical motions since the region dominated by vertical fluctuations reduces to $ 35\%$ of the film's thickness, i.e. horizontal fluctuations dominate over 75\% of the film. From these results, it is apparent that $\lambda$ plays a major role in unbalancing the system and changing the architecture of the hydrodynamic fluctuations so that the confined \textit{Active Carpet} may promote, for example, biogenic mass reorganization mechanisms such as aggregation, recently investigated in a semi-infinite fluid \cite{aguayo2023floating}.

\Cref{Fig:Fig3}(b) illustrates the relationship between $z_{\star}$ and $\lambda$ within the range $\lambda\in[0,4]$, considering different film thicknesses ($20\leq \mathrm{H}\leq 100$). Notably, the graph reveals that $z_{\star}$ decreases with increasing $\lambda$, indicating that stiffer fluid-fluid interfaces promote the expansion of Region III, where horizontal fluctuations outweigh vertical ones. This behavior mirrors the viscous control observed in analogous scenarios, such as superconfined subsurface faults, where viscosity drives fluid flow fluctuations towards quasi-two-dimensional states \cite{ulloa2022energetics}.
Furthermore, our findings indicate that variations in $\mathrm{H}$ exert a more pronounced effect on $z_{\star}$ at lower $\lambda$ values. Specifically, in thin water films ($\mathrm{H}=20$), $z_{\star}$ exhibits weak sensitivity to changes in the strength of the fluid-fluid interface characterized by $\lambda$. Conversely, as the film thickness increases, $z_{\star}$ becomes significantly more responsive to variations in $\lambda$.

\Cref{Fig:Fig3}(c) serves as a comprehensive summary of our findings, encapsulating the essence of our investigation. This figure illustrates the surface representation of $z_{\star}$, depicting its dependency on two pivotal factors: the geometrical confinement denoted by $\rm H$ and the viscous confinement quantified by the ratio $\lambda$. By integrating these parameters, the results effectively delineate the complex dynamics of hydrodynamic fluctuations across different regimes (I, II, III) within the film. It portrays the nonlinear relationship between $z_{\star}$ and the confinement parameters, offering valuable insights into the underlying mechanisms at play. The results in \Cref{Fig:Fig3} underscore the tangible influence exerted by both geometrical and viscous confinements on the spatial structure of velocity fluctuations. This influence extends beyond mere fluctuations, wielding significant control over \textit{Active Carpets} and shaping the trajectories followed by passive tracers. Consequently, an intriguing question arises: Could confined \textit{Active Carpets} potentially catalyze the emergence of aggregated large-scale coherent flow patterns? To delve deeper into this quest, we next examine the coherence of fluid motions and patterns in space by means of pair correlation analysis, mean velocity, and vorticity fields. 

\subsection{Flow coherence and roll-like formation}\label{sec:4c}

\begin{figure*}
\includegraphics[width=1\textwidth]{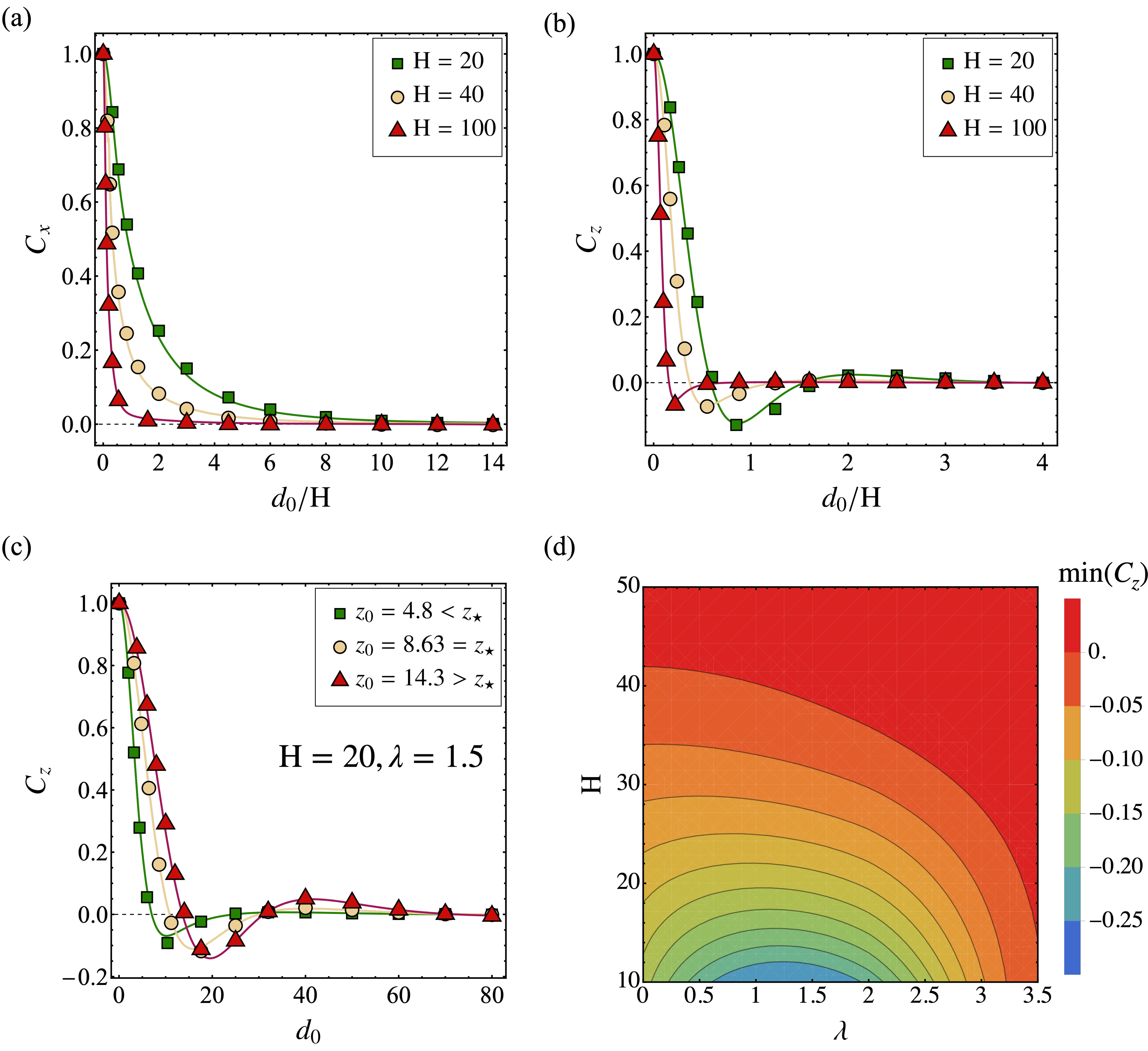}	
    \caption{
    Coherence of flows driven by \textit{Active Carpets}.
    (a, b) Velocity pair correlation function to the relative distance between a pair of tracer particles, $d_0$, normalized by the confinement size relative to the Active Carpet, for $\lambda=1.5,\sigma=1$, and $\kappa=-30$. Markers are simulation points. (c) Vertical velocity pair correlation function, for $\mathrm{H}=20$ and $\lambda=1.5$ on $z_0$ values according to fluctuations regions. Markers are simulation points and solid lines correspond to semi-analytical solutions. (d) Phase diagram of vertical Correlation minima versus $\lambda$ and $\mathrm{H}$.}
\label{Fig:Fig4}
\end{figure*}

A defining feature of living systems is their ability to induce mechanical disturbances on their surroundings over scales greater than their characteristic individual size, prompting investigations into their potential for enhancing mixing in aquatic environments. This capacity of swimming organisms in stirring fluids, particularly through interactions with each other, has garnered significant attention in recent years \cite{sommer2017bacteria,houghton2018vertically,simoncelli2018biogenic,houghton2019alleviation,sepulveda2021persistence,singh2021bacterial,fernandez2022intense}. In the case of microbes swimming in layered, stratified waters--characterized by sharp gradients in density--theoretical arguments suggest individual swimmers have low capacity to mix fluid \cite{more2023motion,wagner2014mixing}. Yet, whether the collective action of small swimming organisms can significantly affect the functioning of the fluid environment remains an open debate \cite{wagner2014mixing,simoncelli2017can,kunze2019biologically}. This motivates our quest for collective effects on large-scale hydrodynamic structures driven by \textit{Active Carpet} within confined layered aquatic systems. Pair velocity correlation emerges as a metric for understanding the relationships between fluid parcel dynamics and biogenic fluid perturbations \cite{belan2019pair,mathijssen2018nutrient,aguayo2023floating}.

We compute the spatial pair velocity correlation to examine how the geometrical and viscous confinement affects the coherence of the flow fluctuations driven by an \textit{Active Carpet}, defined as
\begin{equation}
\label{eq:Corr}
    C_i(d_0) = \dfrac{\langle v_i(\boldsymbol{r}_1) v_i(\boldsymbol{r}_2)\rangle_{e}}{\sqrt{\langle v_{i}^2 (\boldsymbol{r}_1) \rangle_{e} \langle v_{i}^2 (\boldsymbol{r}_2) \rangle}_{e}},
\end{equation}
where $d_0 = | \boldsymbol{r}_2 - \boldsymbol{r}_1 |$ is the Euclidean distance between two fluid parcels in position $\boldsymbol{r}_1$ and $\boldsymbol{r}_2$ with fixed height $z_0$, above the \textit{Active Carpet}. The operator $\langle \cdot \rangle_{e}$ represents an \textit{Active Carpet} ensemble average, and the variance tensor \cref{eq:variance} is utilized to compute the numerator in \cref{eq:Corr}. We performed numerical simulations to compute \cref{eq:Corr} following the next steps. First, we evaluate the flow induced by the \textit{Active Carpet} at $\boldsymbol{r}_{1,2}= (\pm d_0/\sqrt{2},\pm d_0/\sqrt{2},z_0)$ with $z_0=10$ for a wide range of distances $d_0 \in [0.01,200]$. The flow is determined from \cref{eq:total_v} at positions $\boldsymbol{r}_{1,2}$; it results from a single \textit{Active Carpet} ensemble of length $L=500$, using the far-field approximation for the flow field and $N=10^5$ microswimmers, following \citet{guzman2021active}. Then, we compute \cref{eq:Corr} for each ensemble and take the average over ensembles. Results are shown and discussed next. 

\Cref{Fig:Fig4}(a) displays the pair correlation, $C_x$, in the $\hat{e}_x$ direction, representing horizontal correlation, as a function of $d_{0}/{\rm H}$. For simplicity, we omit results for $\hat{e}_y$, which is symmetric to those in $\hat{e}_x$. We examine the horizontal velocity pair correlation between particles separated by a distance $d_{0}$ for three $\mathrm{H}$ values ($\mathrm{H} \in [20,40,100]$), with a fixed $\lambda$ ($\lambda=1.5$) and a representative height $z_0=10$ above the Active Carpet. It is important to note that we normalize $d_0$ to the relative size of the film, $\mathrm{H}$, to the \textit{Active Carpet} to gauge the characteristic length of decorrelation relative to the film size. The varying $\rm H$ range allows us to observe that increasing geometrical confinement (i.e., thinner $\rm H$) expands the relative extent of positively correlated tracer motions over distance $d_0$. In simpler terms, the results indicate that horizontal flows exhibit slower decorrelation with increasing distance $d_0$ in more confined systems. This trend is evident when comparing the curves associated with high geometrical confinement ($\mathrm{H}=20$) and less confinement ($\mathrm{H}=100$). This suggests that for an Active Carpet, disrupting fluid parcels parallel to it is challenging. This phenomenon may be attributed to coherent flows aligning with the constrained motion of microswimmers, resulting in biased, non-prominent horizontal flows. As a notable aside, our numerical findings align with the semi-analytical solution for $\langle v_i(\boldsymbol{r}_1) v_i(\boldsymbol{r}_2)\rangle= \int v_i(\boldsymbol{r}_1) v_i(\boldsymbol{r}_2) Fd\boldsymbol{r}_sd\boldsymbol{p_s}$, shown by solid lines in \Cref{Fig:Fig4}.

Given that the confinement of \textit{Active Carpets} primarily occurs in the $\hat{e}_z$ direction, we delve deeper into the horizontal pair correlation in the $\hat{e}_z$ direction, $C_z$, which we refer to as vertical pair correlation. The results, depicted in \Cref{Fig:Fig4}(b) for $\lambda=1.5$, not only demonstrate good agreement between numerically obtained results (markers) and semi-analytical solutions (lines) but also reveal notable differences from horizontal pair correlation ($C_x$), as shown in \Cref{Fig:Fig4}(a).

An intriguing observation is the significantly shorter decorrelation distance for flow fluctuations in the vertical direction compared to horizontal fluctuations. This is evident from the distance $d_{0}/\left({\rm H}\right)$ over which the $C_{z}$ and $C_{x}$ curves decay to zero. The vertical pair correlation decays at least three times faster than the horizontal pair correlation, suggesting a distinct behavior. Moreover, a key difference is the occurrence of negative values in $C_z$, indicating flow fluctuations in opposite directions at specific distances $d_0$. Such behavior is commonly associated with vortical flow structures, also observed in microswimmer's sheets, active turbulence experiments, and bacterial swarming \cite{tamayo2024swarming,bardfalvy2023collective,mondal2021strong,bardfalvy2023collective}.

In the most confined scenario examined here (see green squares in \Cref{Fig:Fig4}(b)), the curve reaches its minimum $\text{min}(C_z)$ almost at $d_0\approx \mathrm{H}$, suggesting that vortex-like structures scale with the size of the geometrical confinement $\rm H$. The emergence of such large-scale flow structures raises intriguing possibilities. For instance, microbial colonies formed by a monolayer of swimmers, resembling an Active Carpet, could potentially create coherent flow structures in the water column nearly 20 times their characteristic thickness under strong confinement.

In \Cref{sec:Active Carpet in a layered aquatic environment}, we defined the region where active fluctuations exhibit a preferred direction. As this effect originates from the flow itself, we can thoroughly examine the vertical pair correlation $C_z$ in detail. To illustrate how $C_{z}$ changes at different heights, we fix the values of $\mathrm{H}=40$ and $\lambda=1.5$ while varying the position $z_{0}$ at which flows are measured. The results presented in \Cref{Fig:Fig4}(c) demonstrate the sensitivity of $C_{z}$ with respect to height above the \textit{Active Carpet}.
Firstly, notice we measure $C_z$ as a function of $d_0$. Interestingly, we observe similar trends to those depicted in \Cref{Fig:Fig4}(b), where the correlation reaches its minimum at a distance proportional to the size of confinement. This suggests that vertical fluctuations dominate within this region. Notably, the curves exhibit more fluctuations compared to the previous case, which could be attributed to increased confinement. This interpretation aligns with our previous observations.

We have highlighted that the vertical pair correlation reaches a minimum, indicating a region where fluid experiences vertical flow fluctuations in opposing directions. To further explore the intensity of this negative correlation, we vary $\mathrm{H}$ and $\lambda$. \Cref{Fig:Fig4}(d) presents the minimum vertical pair correlation $\text{min}(C_{z})$ for $\mathrm{H}\in[10,50]$ and $\lambda\in[0,3.5]$, computed using the semi-analytical approach. Our analysis reveals that the negative correlation, and thus the intensity of opposing flows, becomes stronger with increased geometrical confinement. Interestingly, we also observe that viscosity confinement, controlled by $\lambda$, enhances $\text{min}(C_{z})$ within the range $\lambda \in [0.5,2]$. The nonmonotonic dependence with $\lambda$ remains poorly understood at present, and this warrants future research for extreme conditions. In the limit of a non-slip boundary at the viscosity interface, i.e. $\lambda\rightarrow \infty$, we observe the pair correlations briefly cross the $y$-axis in the planar and vertical directions. The latter suggests the appearance of a weak complex flow structure related to the action of the \textit{Active Carpet} and confinement. Conversely, in the case of a free-surface and unconfined aquatic environment, i.e.  $\lambda\rightarrow 0$ and $H\rightarrow \infty$, the system does not show significant evidence of an emerging fluid flow structure. For a better understanding of the fluid dynamics of this asymptotic scenario, we refer the reader to \citet{fortune2024biophysical}.

\begin{figure*}
    \centering 
\includegraphics[width=1\textwidth]{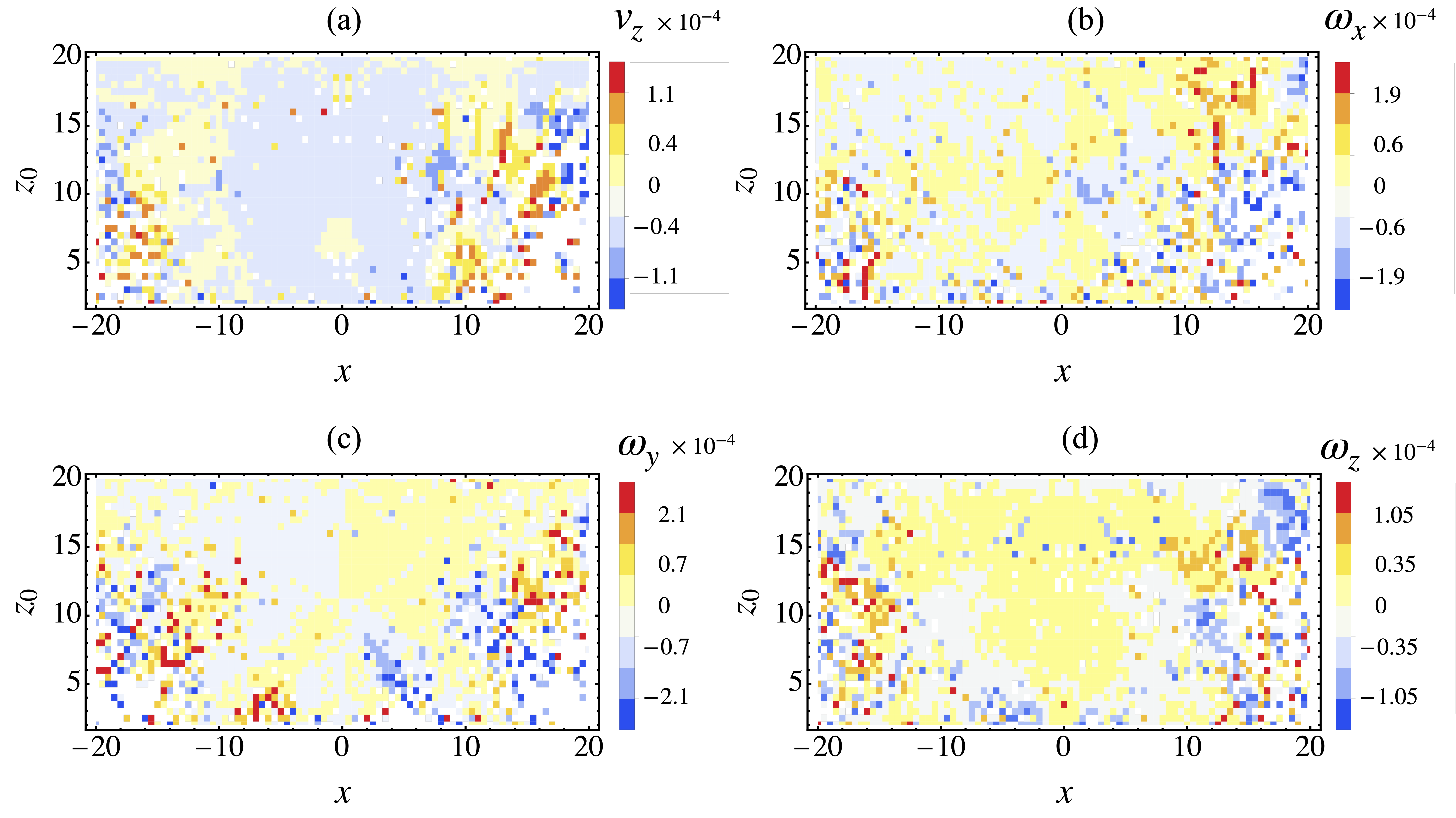}	
    \caption{
    \textit{Active Carpets} can drive large scale recirculations. (a) Average vertical velocity exerted by an \textit{Active Carpet} on fluid parcels across the confinement space, for $\mathrm{H}=20$, $\lambda=1.5$, $\kappa=-30$, and $\sigma=1$. (b,c,d) Vorticity field above the \textit{Active Carpet} where Anti-clockwise (blue) and clockwise (yellow) vortical flows extend over space.} 
\label{Fig:Fig5}%
\end{figure*}

The question of whether confined fluid environments give rise to large-scale flow structures remains unanswered. In our study, we delve into the emergence of vortex-like patterns induced by confined \textit{Active Carpets} by examining the spatial structure of the vertical velocity component and the flow vorticity. We investigate these quantities in a Cartesian coordinate system ($x$, $y$, $z$) to ensure clarity and guide our search for large-scale patterns amidst the biogenic flow fluctuations. In \Cref{Fig:Fig5}(a), we illustrate the velocity field $v_z$ on the $z$-$x$ plane, computed from \cref{eq:flow} using a semi-analytical approach with a spatial resolution of $\Delta z=0.5$ and $\Delta x=0.5$. To highlight coherent trends, we show the sign of the vertical velocity component $v_z$, coloring upward motions in red (+1), downward motions in blue (-1), and no motion in white (0). The results reveal a consistent pattern, demonstrating coherence changes in the vertical flow direction with respect to $x$. Notably, the characteristic length of this `spatial wave pattern' scales as $\rm{H}$. In \Cref{Fig:Fig5}(b,c,d), we show the vorticity components $\omega_x$,~$\omega_y$,~$\omega_z$ on the $x$-$z$ plane. Like panel (a), we represent the sign of $\omega_{i}$ to filter variability and identify large-scale structures and regions with equal vorticity direction. The findings confirm the existence of vortical flow patterns analogous to roll-like formations (exhibiting positive and negative vorticity), with their characteristic vortex length scaling with $\rm H$, especially for $\omega_z$ and $\omega_y$. Our results offer a glimpse into the ability of \textit{Active Carpets} to drive large-scale flow structures in confined environments, bearing striking similarities to phenomena like bio-convection \cite{sommer2017bacteria,thery2020self}, biogenic flows in extremely confined systems \cite{mondal2021strong}, thermally-driven convection in superconfined environments \cite{akashi2019transition,noto2023reconstructing}, and kitchen flows \cite{mathijssen2023culinary}. Such confined flows, characterized by roll-like patterns, are renowned for their high mixing efficiency \cite{sommer2017bacteria,ulloa2022energetics}. In essence, our findings raise fundamental questions about the power of \textit{Active Carpets} inhabiting layered environments to shape their surroundings, enhance the mixing of suspended and dissolved mass in aquatic environments, and facilitate transport between fluid-fluid interfaces and layers.

\section{Conclusions}

We have delved into the hydrodynamics of confined \textit{Active Carpets} within layered aquatic systems. Our study is centered on three main aspects: (i) characterizing the spatial distribution of hydrodynamic fluctuations and the dispersion of passive tracers (\Cref{sec:4a}), (ii) assessing the influence of geometrical and viscous confinement on the hydrodynamics propelled by \textit{Active Carpets} (\Cref{sec:4b}), and (iii) exploring the emergence of macroscopic hydrodynamic structures, referred to herein as roll-like formations (\Cref{sec:4c}).
Our main findings read as follow:
\begin{enumerate}
    \item We derive and report analytical expressions for the active flow fluctuations induced by \textit{Active Carpets} confined in a layer of thickness $\rm H$, between a free surface and a fluid-fluid interface characterized by a viscosity ratio. Our solutions reveal a pronounced non-monotonic behavior in the biogenic hydrodynamic fluctuations, particularly in the vertical direction. This characteristic contrasts with observations made in an \textit{Active Carpet} near a single surface system \cite{guzman2021active,aguayo2023floating}. The non-monotonic dynamics of these hydrodynamic fluctuations are intricately influenced by the degree of geometrical and viscous confinement imposed by $\rm H$ and the viscosity ratio $\lambda = \mu_{2}/\mu_{1}$ at the fluid-fluid interface, where $\mu_{1}$ is the viscosity of the fluid where the \textit{Active Carpet} lives, and $\mu_{2}$ is the viscosity of the deeper layer. Numerical simulations corroborate the findings obtained from analytical solutions. These simulations provide a deeper exploration into the dynamics of passive tracer particles, showcasing the height-dependent topology of motions above the \textit{Active Carpet}.
    
    \item We show how the topology of hydrodynamic fluctuations undergoes transformation depending on the degree of geometrical confinement and the intensity of the viscosity ratio at the fluid-fluid interface boundary. Notably, we have uncovered the presence of three distinct spatial regions: (I) A proximal region to the \textit{Active Carpet} where vertical fluctuations supersede horizontal fluctuations; (II) An intermediate region characterized by isotropic hydrodynamic fluctuations; and (III) The furthest region from the \textit{Active Carpet} where horizontal fluctuations prevail over vertical fluctuations. This newfound understanding equips us with a tool to wield control over directed motion within confined vertical spaces. Furthermore, it illustrates how both dependencies are intrinsically linked to the amplification or diminishment of regions of agitation.
    
    \item By measuring the velocity pair correlation of the exerted hydrodynamic fluctuations induced by the confined \textit{Active Carpet}, we demonstrate the existence of coherent vortical motion—predominantly propelled by vertical flows. Our investigation pinpointed that the characteristic length of the formed roll-like patterns is intimately tied to the thickness of the confined layer harboring the \textit{Active Carpet}. Remarkably, these coherent vortical structures manifest with exceptional prominence in highly confined systems and in the presence of sharp viscosity jumps at the fluid-fluid interface. 

\end{enumerate}

These findings carry implications for our comprehension of microbial swimmers that flourish and cultivate biofilms at interfaces in natural shallow water environments \cite{desai2020biofilms}, such as (i) ponds etched upon soil and ice \cite{mohit2017hidden,anesio2017microbiome}, (ii) shallow saline lagoons and wetlands \cite{de2014heat}, (iii) streams \cite{battin2016ecology}, inside the human body \cite{maheshwari2019colloidal, figueroa2019mechanical} and even in human habitats like the thin films of water found in kitchens, bathrooms, swimming pools and laboratories \cite{flores2013diversity,novak2020microorganisms}.

By shedding light on the dynamics of these active flows, our research offers insights into the behavior of microbial communities in thin layers within aquatic ecosystems. From pristine natural settings to human-altered landscapes, the understanding of the hydrodynamics induced by \textit{Active Carpets} is crucial for devising strategies to manage and harness the potential of microbial populations for various applications, ranging from water remediation to microfluidics and biotechnology.

\section*{Acknowledgment}
F.A.B., G.A. and F.G.-L. have received support from the ANID – Millennium Science Initiative Program – NCN19 170, Chile. F.G.-L. was supported by Fondecyt Iniciación No.\ 11220683. H.N.U. and A.J.T.M.M. were supported by start-up grants from the University of Pennsylvania.
A.J.T.M.M. acknowledges funding from the United States Department of Agriculture (USDA-NIFA AFRI grants 2020-67017-30776 and 2020-67015-32330), the Charles E. Kaufman Foundation (Early Investigator Research Award KA2022-129523) and the University of Pennsylvania (University Research Foundation Grant and Klein Family Social Justice Award).

%

\end{document}